\documentclass[final,3p,times]{elsarticle}
\usepackage{graphicx}
\usepackage{amsmath}
\usepackage{amssymb}
\usepackage{xcolor,soul}
\sethlcolor{yellow}

\usepackage{multirow}

\usepackage{setspace}

\usepackage{hyperref}
\usepackage{color}
\usepackage{epstopdf}
\usepackage{tabularx}
\usepackage{subfig}
\usepackage[font=small,labelfont=bf]{caption}
\usepackage{amssymb}
\usepackage{amsmath}
\usepackage{multirow}
\usepackage{array}

\begin{document}
	\doublespacing
	\begin{frontmatter}

\title{Buckling and yield strength estimation of architected materials under arbitrary loads}
 
\author[add1]{Morten N. Andersen \corref{cor1}}
\author[add2]{Yiqiang Wang} 
     \author[add1]{Fengwen Wang}
     \author[add1]{Ole Sigmund}  
   \address[add1]{Department of Mechanical Engineering, Solid Mechanics,  
    Technical University of Denmark, DK-2800, Lyngby, Denmark} 					  
    \address[add2]{State Key Laboratory of Structural Analysis for Industrial Equipment, 
Dalian University of Technology, 116024, China}
\cortext[cor1]{email: morand@mek.dtu.dk; Tel. :+45 4525 4209;  Fax: +45 4593 1475.}

\begin{abstract}
Buckling strength estimation of architected materials has mainly been restricted  to load cases oriented along symmetry axes. However, realistic load scenarios normally exhibit more general stress distributions. In this paper we propose a simple yet accurate method to estimate the buckling strength of stretch-dominated lattice structures based on individual member analysis. As an integral part of the method, the yield strength is also determined. This simplified model is verified by rigorous numerical analysis. In particular, we efficiently compute the complete buckling strength surfaces of an orthotropic bulk modulus optimal plate lattice structure and isotropic stiffness optimal plate and truss lattice structures subjected to rotated uni-axial loads, where the ratio between the highest and lowest buckling strength is found to be $1.77$, $2.11$ and $2.41$, respectively. For comparison, we also provide their yield strength surfaces, where the corresponding ratios are $1.84$, $1.16$ and $1.79$. Furthermore, we use the knowledge gained from the simplified model to create a new configuration of the isotropic plate lattice structure with a more isotropic buckling strength surface and buckling strength ratio of $1.24$, without deterioration of the stiffness or yield strength. The proposed method provides a valuable tool to quickly estimate the microstructural buckling strength of stretch-dominated lattice structures, especially for applications where the stress state is non-uniform such as infill in additive manufacturing.
\end{abstract}


\end{frontmatter}

\section{Introduction} \label{sec:intro}
Architected materials with ultimate stiffness and strength are pushing the limits for high performance light-weight structures. In particular, exploiting architected materials as infill for additive manufacturing has shown to vastly increase macroscopic buckling strength while decreasing stiffness very little compared to equal volume solid counterparts. For instance, Clausen et. al. \cite{clausen2016exploiting} obtained over $400\%$ buckling strength improvement, while only losing $20\%$ in stiffness for a particular beam structure. However, with ever increasing demands for light-weight, optimized structures and advances in manufacturing techniques, plate and truss microstructures are realizable with lower volume fractions and higher slenderness, which make them prone to local buckling failure and therefore necessitate efficient microscopic stability analysis.

The classical approach to studying microstructural buckling is to analyze the buckling strength of individual plate or truss members, where the geometry and boundary conditions are approximated as known analytical solutions  \cite{christensen1986mechanics,valdevit2013compressive,latture2018design,tancogne2018elastically}. Deshpande \cite{deshpande2001effective} concluded that simply supported boundary conditions may only be viewed as lower bounds, which implies that clamped boundary conditions may similarly be viewed as upper bounds. On the other hand, the microstructural buckling strength of various simple 2D honeycomb structures have been analytically studied by using Floquet-Bloch wave theory which is capable of capturing both local and global buckling modes \cite{ohno2004long,haghpanah2014buckling}. Furthermore, microstructural buckling of both random and periodic porous elastomers under large deformations have been investigated numerically \cite{triantafyllidis2006failure,michel2007microscopic}. In 3D, buckling failure of random porous elastomers has been investigated by using second order homogenization theory which assumes linear comparison composites, and the macroscopic instability is detected by loss of strong ellipticity in the homogenized constitutive model \cite{lopez2007homogenization1,lopez2007homogenization2}. Recently, a systematic study on microstructural buckling strength of several 3D lattice structures was performed, where the numerical Floquet-Bloch wave theory combined with expensive FE analysis was applied to determine the critical buckling strength accounting for both local and global buckling \cite{andersen2020competition}.

While the classical approach is simple, it is based on strict assumptions which in reality might not be appropriate.  Furthermore, the joints created by intersection regions could affect the stress field and therefore have an impact on the estimation accuracy of the buckling strength. This effect is amplified in high volume fractions. On the other hand, the more sophisticated models provide higher accuracy but require significantly more development efforts and computational cost. All these studies have mainly focused on the particularly well-defined load cases with the loading directions oriented along the principal axes. However, to the best of our knowledge, no investigation has been carried out for directional buckling strength surfaces of 3D architected materials due to computational cost.  Hence a cheaper buckling strength estimation method is highly desired to systematically investigate the material buckling strength under arbitrary loads.

\begin{figure}[!t]
  \centering
  \includegraphics[width=0.65\textwidth]{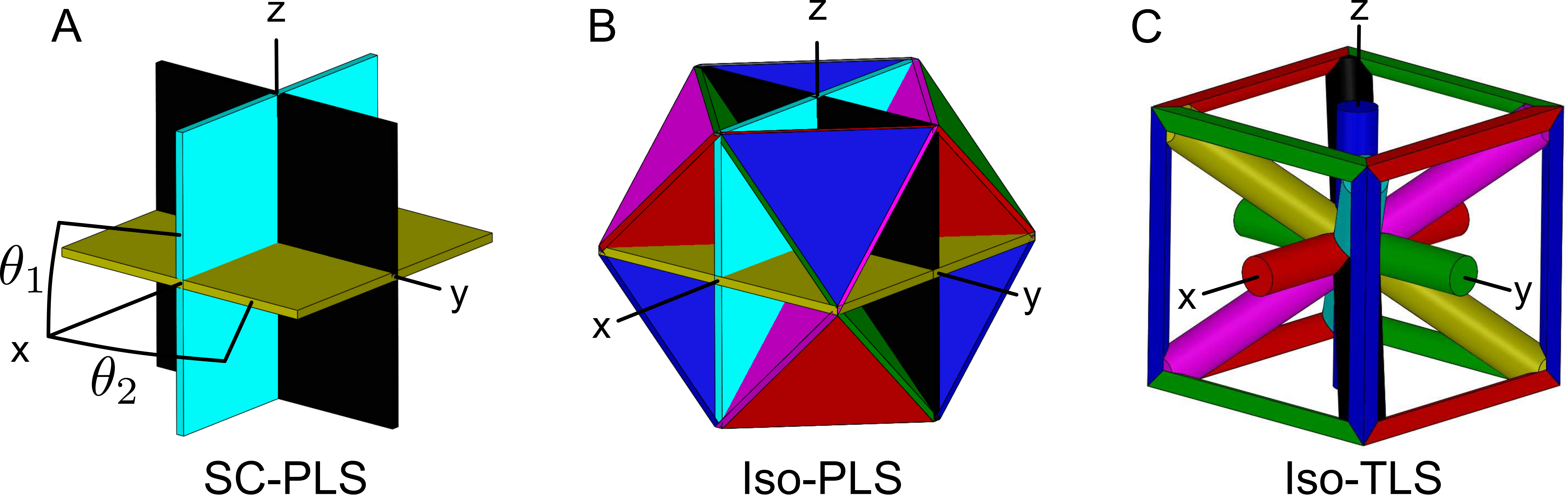}
\caption{Investigated lattice structures. A: Simple cubic plate lattice structure (SC-PLS). B: Simple cubic+body centered cubic isotropic plate lattice structure (Iso-PLS). C: Simple cubic+body-centered cubic isotropic truss lattice structure (Iso-TLS). The members are color coded where members with the same color have the same orientation and therefore constitute a member set.}
\label{fig:lattices}
\end{figure}

In this paper, we propose a simple yet quite accurate method to estimate the elastic buckling strength of plate and truss lattices based on individual member buckling analysis, where the topology of the lattice is accounted for. The proposed simplified method is validated with linear material buckling analysis \cite{andersen2020competition}, which has been shown to provide good predictions compared to nonlinear microstructural buckling studies - at least in 2D \cite{wang2020numerical}. The proposed method  accounts for cell-local instability since this tends to be the critical mode for  materials with sufficient shear stiffness \cite{andersen2020competition}. We apply the method to study the uni-axial buckling strength as a function of the load direction, but the method is general and valid for arbitrary loads. As representatives of different types of lattice structures, we study an orthotropic simple-cubic plate lattice structure (SC-PLS), an isotropic plate lattice structure (Iso-PLS) and an isotropic truss lattice structure (Iso-TLS) from the literature \cite{berger2017mechanical, gurtner2014stiffest} shown in figure \ref{fig:lattices}. Specifically, these isotropic lattice structures are a combination of simple cubic (SC) and body-centered cubic (BCC) members. Interestingly, while the PLSs offer superior stiffness and yield strength, the microstructural yield and buckling strength surfaces can be vastly different. In contrast, the TLS strength surfaces are very similar for both yield and buckling failure, in turn resulting in more predictable failure mechanisms purely based on volume fractions.  Furthermore, based on the knowledge and results obtained from the simplified model, we suggest an Iso-PLS which has an improved isotropic buckling strength surface, with no deterioration of the stiffness and yield strength surfaces. 

The proposed simplified model provides a valuable tool to quickly estimate the buckling strength of lattice structures, especially when the loading is complex such as when used for infill in additive manufacturing. Additionally, the method provides insight into the overall performance by identification of the critical failure members of the lattice structures.

The paper is organized as follows. Section \ref{sec:Method} first briefly summarizes linear material buckling strength analysis  and then present the proposed simplified modelling approach and corresponding validations. Section \ref{sec:Results} employs the proposed method to study the buckling strength of the investigated lattice structures under rotated uni-axial loads and the yield surface. Furthermore, the results and the accuracy of the buckling strength estimations are verified with linear material buckling analysis.  Finally, conclusions are drawn in section \ref{sec:conc}.

%%%%%%%%%%%%%%%%%%%%%%%%%%%%%%%%%%%%%%%%%%%%%%%%%%%%%%%%%%%%%
\section{Method} \label{sec:Method}
Buckling modes of lattice structures can be local, global, or in-between \cite{andersen2020competition}, where triggering of each mode type depends on the local stress state in the members. Local modes are characterized as periodic or anti-periodic cell-wise buckling modes i.e. the periodicity of the buckling mode span one or two unit cells. Global modes are characterized by the periodicity of buckling modes being much larger than the size of the unit cell. Although, deeper investigations are needed to definitively determine what triggers the global modes, a general promoter of global (shear) modes is related to low shear stiffness, which can be seen from the band diagram presented in \cite{andersen2020competition}. Note that, the Zener anisotropy ratio \cite{zener1949elasticity} for local mode dominated lattice structures such as SC-PLS, Iso-PLS, and Iso-TLS are $0.45$, $1.00$ and $1.00$, respectively, where a ratio of $1.00$ means isotropy. In contrast, the SC-TLS from \cite{andersen2020competition}, has a Zener anisotropy ratio of $0.12$ and exhibits a global critical buckling mode. Assuming sufficient shear stiffness, buckling instabilities in stretch-dominated lattice structures are thus predominantly governed by cell-local modes. That is, the plate and truss components buckle in repeated periodic or anti-periodic cell-wise patterns. Hence, it is possible to estimate the microstructural buckling strength by performing analysis on individual plate or truss members, where the lowest member strength determines the effective buckling strength. 

In the following section, we first briefly summarize the linear material buckling analysis in \cite{andersen2020competition} and then focus on our proposed simplified model using member buckling analysis and validations.

\subsection{Linear material buckling strength analysis}\label{sec:FloquetBloch}
The material buckling strength of a lattice structure subjected to a global stress state $\boldsymbol{\sigma}^0=\left[\sigma_1, \sigma_2, \sigma_3,\sigma_4, \sigma_5, \sigma_6 \right]^T$ is calculated by linear material buckling analysis using the finite element method. First, the homogenized elasticity matrix, $\boldsymbol{D}^H $ , is calculated using a computational homogenization method \cite{guedes1990preprocessing,sigmund1994materials,bensoussan2011asymptotic}. The global stress state $\boldsymbol{\sigma}^0$ is converted to a global strain $\boldsymbol{\varepsilon}^0$ and then to a local pre-stress $\boldsymbol{\sigma}^e$ in a given element $e$, stated as 
\begin{align}
	\boldsymbol{\sigma}_e = \boldsymbol{D}_e  \boldsymbol{\varepsilon}^0  =\boldsymbol{D}_e \left(\boldsymbol{D}^H \right)^{-1} \boldsymbol{\sigma}^0,
\end{align}
where $\boldsymbol{D}_e$ is the elasticity matrix of the material in element $e$. A subsequent  linear  buckling analysis is performed to calculate the material buckling strength, where instabilities on the short- and long wavelength are captured by employing Floquet-Bloch boundary conditions \cite{triantafyllidis1993comparison,neves2002topology,thomsen2018buckling,andersen2020competition}, given as
\begin{align}\label{eq:buckling}
&\left[  \boldsymbol{K}_0 +   \lambda  \boldsymbol{K}_{\sigma}  \right]\boldsymbol{\Phi} = \boldsymbol{0} \\
  \boldsymbol{\Phi}|_{x=\Gamma_\text{Front}}=e^{ik_1}\boldsymbol{\Phi}|_{x=\Gamma_\text{Back}},  \quad
&\boldsymbol{\Phi}|_{y=\Gamma_\text{Right}}=e^{ik_2}\boldsymbol{\Phi}|_{y=\Gamma_\text{Left}},\quad  \boldsymbol{\Phi}|_{z=\Gamma_\text{Top}}=e^{ik_3}\boldsymbol{\Phi}|_{z=\Gamma_\text{Bottom}}. \nonumber
\end{align}	
where the material unit cell has unit size,  $\boldsymbol{K}_0$ is the initial stiffness matrix, $\boldsymbol{K}_\sigma$ is the stress stiffening matrix and $\lambda$ is the eigenvalue with the associated eigenvector $\boldsymbol{\Phi}$. The eigenvector is supplemented with the Bloch function $e^{i\boldsymbol{k}}$, where $i=\sqrt{-1}$ is the imaginary unit, $\Gamma_\square$ are the faces of the unit cell and $\boldsymbol{k}=\left[k_1, k_2, k_3 \right]^T$ is the wavevector, which modifies the boundary conditions of the unit cell, enabling capturing of local and global buckling modes. The material buckling strength is defined as $\sigma_c=\lambda_\text{min} \| \boldsymbol{\sigma}^0 \|$, where $\| \cdot \|$ represents the $2$-norm and $\lambda_\text{min}$ is the lowest eigenvalue over all wavevectors $\boldsymbol{k}$. The lowest eigenvalue is determined by sweeping $\boldsymbol{k}$ along the irreducible Brillouin zone edges.  The irreducible Brillouin zone is determined by shared symmetries in the geometry and loading \cite{andersen2020competition}. However, for complex load cases without load symmetries, the required sweep becomes significantly larger and may cover the entire Brillouin zone. Even for load cases that can be reduced to the edges of a subpart of the irreducible Brillouin zone, a large number of 3D linear buckling analyses (Eq.~\eqref{eq:buckling}) is required to accurately determine the material buckling strength (CPU times up to $100$ hours on supercomputer as discussed at the end of section \ref{sec:Results}). Hence, a cheaper buckling strength estimation method is highly desired to systematically investigate the material buckling strength under arbitrary loads.

\subsection{The simplified model}
Linear material buckling strength analysis in  \cite{andersen2020competition} showed that the critical buckling modes of materials with sufficient shear stiffness are dominated by local member buckling. Hence material buckling strength can be estimated based on local member buckling analysis. In this section, we propose a simplified model to estimate material buckling strength, which is a compromise between overly simplified analytical calculations \cite{christensen1986mechanics,valdevit2013compressive,latture2018design,tancogne2018elastically,deshpande2001effective} and expensive linear material buckling strength analysis described in the previous subsection. In the proposed simplified model, we estimate the material buckling strength of the lattice structures based on the buckling analysis of individual members, where the unit cell topology is considered. Adjacent members that provide rotational stiffness to the boundaries are approximated and converted to moments that still provide rotational stiffness but also reduce the geometric complexity. Before the simplified model is presented, the local member stress calculation is described.

%In the proposed simplified model, we consider the unit cell topology and the adjacent members contribute to local member buckling analysis as stiffeners. The rotational stiffness from adjacent members is approximated and converted to moments which still provide rotational stiffness but also reduces the geometric complexity. \textcolor{red}{Before the simplified model is presented, the local member stress calculation is described.}

\subsubsection{Stress calculation in local members}\label{ch:StressCalc} 
The global stress $\boldsymbol{\sigma}^0$ is converted to the local stress state in each truss or plate member using the homogenized elasticity matrix $\boldsymbol{D}^H$ \cite{christensen1986mechanics,bourdin2008optimization}. The global stress is first converted to global strain via 
\begin{equation} \label{eq:GlobStrain}
\boldsymbol{\varepsilon}^0=\left(\boldsymbol{D}^H \right)^{-1} \boldsymbol{\sigma}^0.
\end{equation}
The global strain (converted to a matrix as indicated by the subscript $\text{m}$) is then projected to each local member coordinate system using

\begin{equation}\label{eq:RotStrain}
\boldsymbol{\varepsilon}_\text{m}'=\mathbf{T}\boldsymbol{\varepsilon}_\text{m}^0 \mathbf{T}^T
\end{equation}
where $\mathbf{T}$ is a rotation matrix describing the local  member coordinate system. Finally, the projected strain  is converted to the local member stress by 

\begin{equation}\label{eq:LocStress}
    \boldsymbol{\sigma}_\text{loc}= \boldsymbol{D}_\text{loc} \boldsymbol{\varepsilon}'
\end{equation}
where $\boldsymbol{D}_\text{loc}$ is the local elasticity matrix. 

It is noted that $ \boldsymbol{\sigma}_\text{loc}$ is a stress vector for PLSs and a scalar for TLSs as plates provide planar stiffness while trusses only provide uni-axial stiffness. Note that parallel members in one group (see the colored members in figure \ref{fig:lattices}) are subject to identical local stresses. This assumption is only strictly true in the low volume fraction limit. Nevertheless, this stress approximation approach is also used here for moderate volume fractions for the sake of simplicity. Herein, we use normalized Young's modulus, $E_0=1$ with Poisson's ratio, $\nu_0=1/3$ for the base material.  Based on the stress calculation, the microstructural yield strength is defined as
\begin{equation} \label{eq:YieldStrength}
	\sigma_y=\tilde{\sigma}_y \sigma_0 = \frac{\| \boldsymbol{\sigma}^0 \|}{\sigma_\text{vm,max}}   \sigma_0
\end{equation}
where $\sigma_\text{vm,max}$ is the maximum von Mises stress of all members under the given load case, $\sigma_0$ is the yield strength of the base material and $\tilde{\sigma}_y$ denotes the normalized yield strength. Utilizing normalized strength properties is beneficial as different base materials can be inserted into eq. \eqref{eq:YieldStrength}. 
%Variability in terms of Poisson's ratio is small in the usual range of (compressible) base material values $\nu_0 \in \left[0, 1/2\right[$, and hence choosing a value of $\nu_0=1/3$ yields negligible differences in the stiffness and thus strength in the aforementioned interval \cite{andersen2020competition}.

%%\textcolor{red}{In the following subsection, we present the approach to the simplified model comprising of how the adjacent members provide rotational stiffness to the boundaries with how to approximate and implement the rotational stiffness.}

\subsubsection{Rotational stiffness and buckling analysis}\label{sec:RotStiff}

\begin{figure}[!t]
  \centering
  \includegraphics[width=0.7\textwidth]{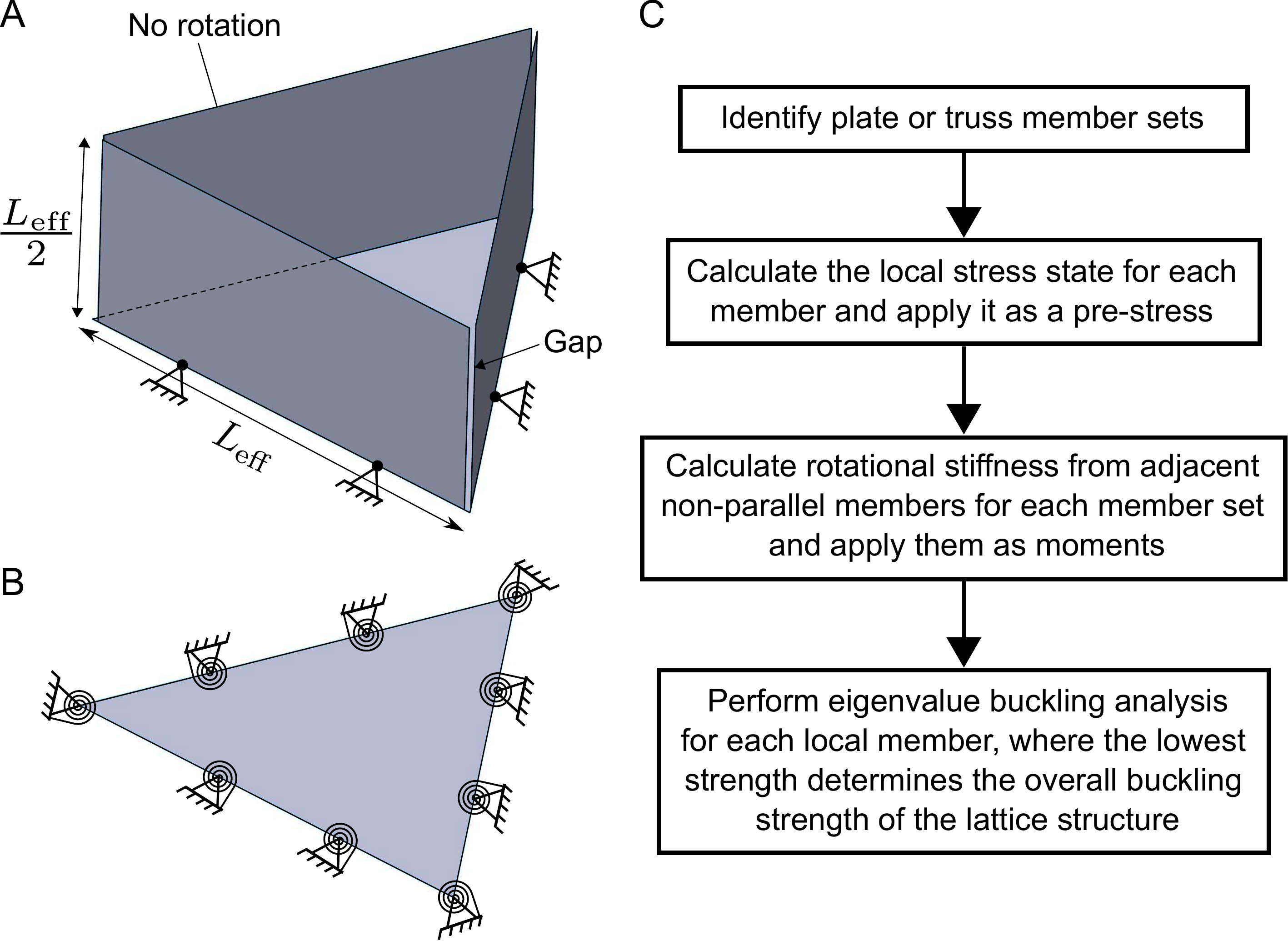}
\caption{Simplified modelling approach. A: Iso-PLS base plate (c.f. one of the colored BCC triangles in figure \ref{fig:lattices}B) with one set of adjacent wall stiffeners (with Poisson's ratio, $\nu_0=0$). The edges of the base plate have simply supported boundary conditions, while the top edges of the wall stiffeners have no rotation allowed and hence mimic a wide cantilever beam with an end moment. B: Resulting base plate with simply supported boundary conditions with the addition of rotational springs. C: Flow chart describing the simplified model approach to microstructural buckling strength estimation.}
\label{fig:Stiffeners}
\end{figure}

The simplest material buckling strength estimations consist of approximating the truss and plate members as known analytical solutions, i.e. simple column or plate buckling with simply supported or clamped boundary conditions. However, as the individual plate or truss members are connected to varying numbers of adjacent members that provide rotational stiffness, neither simply supported nor clamped boundary conditions are appropriate but may only be used as lower or upper bounds. We propose a simple and flexible method based on elastic boundary conditions where the adjacent members determine the rotational stiffness. Several approaches may be envisioned to account for adjacent members, but the proposed method comprises applying simple supports with the addition of rotational springs to the boundaries/ends to approximate the rotational stiffness contributions from neighbor members. Herein, the spring stiffness of adjacent members, i.e. for both plates and trusses, are approximated by beam theory, where plate stiffeners are treated as wide beams (see figure \ref{fig:Stiffeners}). Furthermore, the critical buckling modes of plates and trusses with either simply supported or clamped boundary conditions are assumed to deform in half-sine waves (based on extensive numerical observations). Therefore, the springs are approximated by the stiffness of a cantilever beam subjected to an end moment, as it resembles a quarter sine wave. The rotational stiffness, denoted by $k_t$, can thus be applied as a moment that acts in the opposite direction of the buckling deformation and stiffens the boundaries. The boundary moment $M$ is defined as
\begin{equation}
	M=k_t\phi
\end{equation}
where $\phi$ is the rotation angle around plate edges or truss ends. This equation has the same form as a cantilever beam subjected to an end moment, which is

\begin{equation}
	M=\frac{E_0I}{(L_\text{eff}/2)}\phi
\end{equation}
where $I$ is the second moment of inertia of the adjacent stiffener with $\nu_0=0$ allowing for analytical expression and treating plates as wide beams, $L_\text{eff}$ is the effective length defined as $L_\text{eff}=L-t/2$, where $L$ and $t$ are the member length and thickness, respectively. The effective length is used to account for shortening effects from intersecting members. This assumption is grossly simplified but is used here for the sake of simplicity and keeping identical assumptions for both types of lattice structures. Furthermore, adjacent parallel members do not add rotational stiffness as they are subjected to the same local stress and therefore buckle in a continuous wave. Consequently, the rotational stiffness of a non-parallel adjacent stiffener is split between the connected parallel members. The resulting moment summed up from adjacent members is

\begin{equation}\label{eq:MomentSum}
	M=\sum_{n=1}^{N}\left( \frac{\frac{E_0I_n}{(L_{\text{eff},n}/2)}}{2} \right)\phi = \sum_{n=1}^{N}\left( \frac{E_0I_n}{L_{\text{eff},n}} \right)\phi
\end{equation}
where $N$ indicates the number of adjacent non-parallel members.

Based on Timoshenko beams or Mindlin plates with quadratic finite elements, linear buckling analysis is performed  to determine the buckling strength of the $i$'th member, given as 
\begin{equation}\label{eq:EigBuck}
\left[  \boldsymbol{K}^i_0 +   \lambda_1^i  \boldsymbol{K}^i_{\sigma}  \right]\boldsymbol{\Phi}_1^i = \boldsymbol{0}. 
\end{equation}
where $\lambda_1^i$ and $\Phi_1^i$ are the fundamental eigenvalue and associated buckling mode of the $i$'th member. The material buckling strength is determined by  the minimum eigenvalue among all the members, given as

\begin{equation}\label{eq:BucklingStrength}
	\sigma_c=  \left( \min\limits_i  \lambda_1^i\right)  \  \| \boldsymbol{\sigma}^0 \|  
\end{equation}
where the normalized buckling strength is defined as $\tilde{\sigma}_c = \sigma_c / E_0$. 	
The flow chart of the proposed simplified model is presented in figure \ref{fig:Stiffeners}C. It is summarized in four steps:  
\begin{enumerate}
  \item Identify plate or truss member sets (figure \ref{fig:lattices}).
  \item Calculate the local stress for each member set and apply it as pre-stress (Eqs.  \eqref{eq:GlobStrain}-\eqref{eq:LocStress}).
  \item Calculate rotational stiffness from adjacent non-parallel members for each member set and apply them as moments (Eq. \eqref{eq:MomentSum}).
  \item Perform eigenvalue buckling analysis for each member set. The lowest eigenvalue determines the overall material buckling strength (Eqs. \eqref{eq:EigBuck}-\eqref{eq:BucklingStrength}).
\end{enumerate}

This planar FE analysis is obviously significantly cheaper than a full 3D Floquet-Bloch analysis.

\subsection{Validation}

In this subsection, we first illustrate and verify the proposed simplified model using buckling analysis of a truss and plate member with stiffeners (figure \ref{fig:SM_Ex_IsoTLS} and \ref{fig:SM_ex_SCP}). Then, the simplified model is validated by estimating the material buckling strength versus volume fractions of the investigated lattice structures subjected to a uni-axial load (figure \ref{fig:BC_study}).
\subsubsection{Member buckling}
\begin{figure}[!t]
  \centering
  \includegraphics[width=1\textwidth]{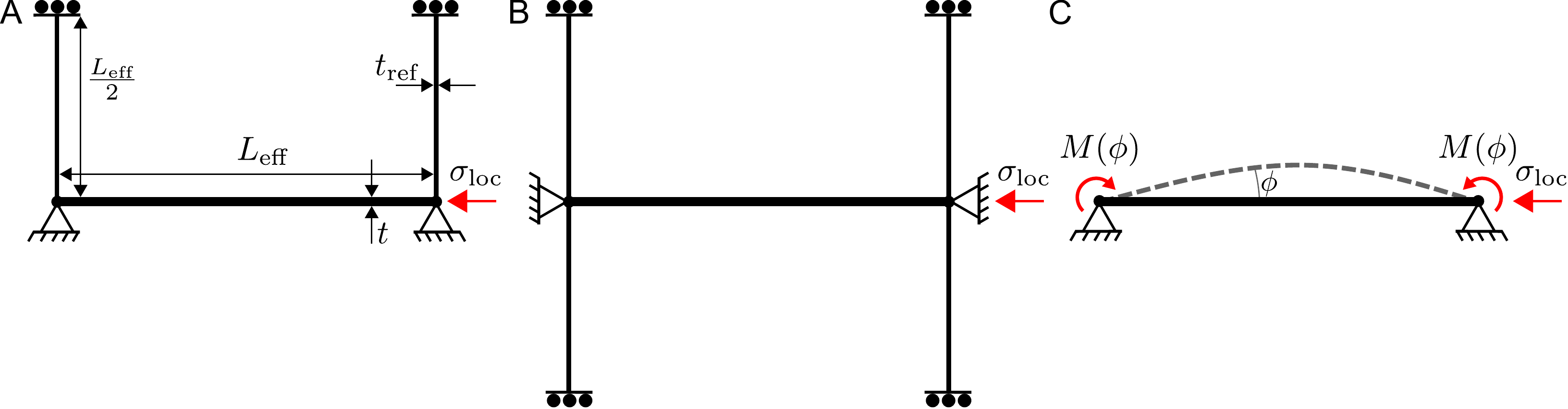}
\caption{Simplified model truss example. The base truss is the thick horizontal truss, which has simply supported boundary conditions at the ends. The thinner vertical trusses are truss stiffeners that have no rotation allowed at the opposite ends of the base truss ends. A: Base truss with one truss stiffener at each end. B: Base truss with two truss stiffeners at each end. C: The corresponding simplified model where the rotational stiffness from adjacent truss stiffeners has been converted to moments.}
\label{fig:SM_Ex_IsoTLS}
\end{figure}

\begin{figure}[!b]
  \hspace*{-0.95cm} 
  \includegraphics[width=1.05\textwidth]{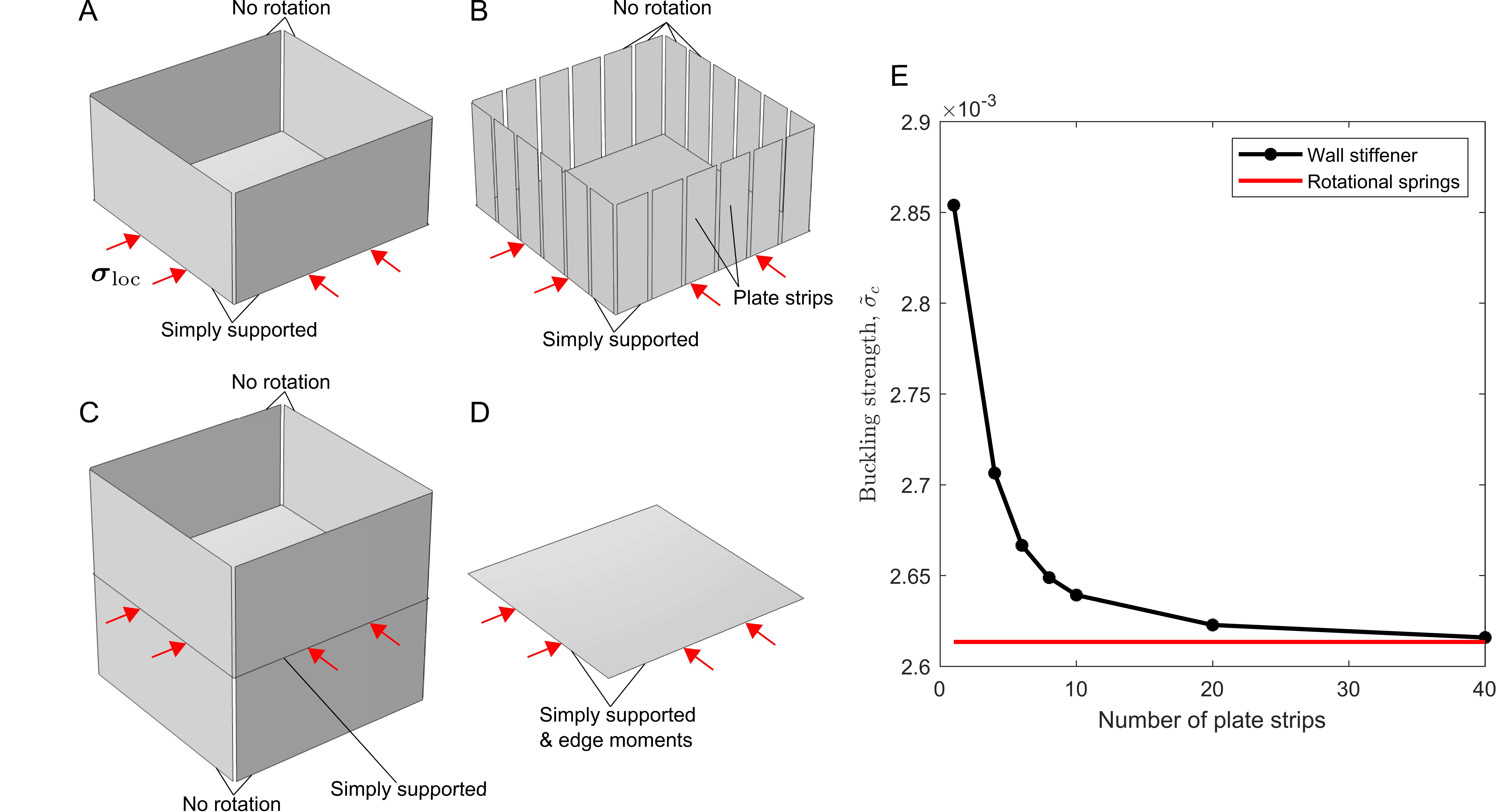}
\caption{Simplified model plate example. The base plate is the horizontal plate, which has simply supported boundary conditions at the edges. The vertical plates are plate stiffeners with $\nu_0=0$ and have no rotation allowed at the opposite edges of the base plate edges. A: Base plate with one plate stiffener per edge. C: Base plate with two plate stiffeners per edge. D: The corresponding simplified model where the rotational stiffness from adjacent plate stiffeners has been converted to moments (Analogous to figure \ref{fig:SM_Ex_IsoTLS}C). B: Base plate with one plate stiffener per edge which has been split into plate strips. E: Normalized buckling strength evaluation versus the number of plate strips. The red curve corresponds to figure \ref{fig:SM_ex_SCP}D and the black curve corresponds to figure \ref{fig:SM_ex_SCP}B.}
\label{fig:SM_ex_SCP}
\end{figure}

A truss and a plate example with stiffeners are considered to illustrate and verify the proposed simplified model.  Specifically, one and two adjacent stiffeners per truss end or plate edge are compared to the corresponding simplified model, where the rotational stiffness from adjacent stiffeners is converted to moments. The truss example is shown in figure \ref{fig:SM_Ex_IsoTLS}, where the plate example is analogous and shown in figure \ref{fig:SM_ex_SCP}. The applied load is aligned with the Cartesian x-axis (see figure \ref{fig:lattices}).
\begin{table}[!b]
\centering
{\renewcommand{\arraystretch}{1.2} %<- modify value to suit your needs

\begin{tabular}{|l|c|c|c|}
\hline
\multicolumn{1}{|c|}{Case} & Rotational springs & Member stiffeners & Error $\%$ \\ \hline
Iso-TLS: 1 stiffener \space (Figure \ref{fig:SM_Ex_IsoTLS}A)      & 9.928E-4           & 9.928E-4         & 0          \\ \hline
Iso-TLS: 2 stiffeners (Figure \ref{fig:SM_Ex_IsoTLS}B)     & 1.196E-3           & 1.196E-3         & 0          \\ \hline
SC-PLS: 1 stiffener \space (Figure \ref{fig:SM_ex_SCP}A)       & 2.613E-3           & 2.854E-3         & 8.44       \\ \hline
SC-PLS: 2 stiffeners (Figure \ref{fig:SM_ex_SCP}C)       & 2.939E-3           & 3.319E-3         & 11.44      \\ \hline
\end{tabular}
}
\caption{Comparison of member stiffener versus rotational springs exerted as moments.}
\label{tab:Stiffener}
\end{table}

The truss example is a simplification of the red SC truss of the Iso-TLS (see figure \ref{fig:lattices}C), however only stiffened by one or two trusses per end. The truss length and thickness are $L=1$ and $t=0.144$, respectively, resulting in an effective length $L_\text{eff}=L-t/2=0.928$. The reference thickness $t_\text{ref}$ of the adjacent stiffener that provides half rotational stiffness to the base SC truss, is calculated from $I(t)/2 = I(t_\text{ref})$ resulting in $t_\text{ref}=t/ \sqrt[4]{2}$. Furthermore, the applied axial stress to the base truss is $\sigma_\text{loc}=-19.292$. Figure \ref{fig:SM_Ex_IsoTLS}A and B show the base truss with one and two truss stiffeners per end, respectively. Figure \ref{fig:SM_ex_SCP}C shows the corresponding simplified model where the adjacent truss stiffeners have been converted to end moments proportional to the end rotation $\phi$. The resulting buckling strengths are shown in table \ref{tab:Stiffener}. The results from truss stiffeners and rotational springs are identical for both one and two adjacent stiffeners. This confirms that the simplification to rotational stiffness proposed here is effective.

The plate example is based on the yellow plate of the SC-PLS (see figure \ref{fig:lattices}A). The plate side length and thickness are $L=1$ and $t=0.0717$, respectively, resulting in an effective length of $L_\text{eff}=L-t/2=0.964$. The reference thickness $t_\text{ref}$ of the adjacent stiffener that yields half rotational stiffness to the base plate, is calculated from $I(t)/2 = I(t_\text{ref})$ resulting in $t_\text{ref}=t/ \sqrt[3]{2}$. Furthermore, the planar stress applied to the base plate is $\boldsymbol{\sigma}_\text{loc} = \left[-7.168, -1,230, 0 \right]^T$. Figures \ref{fig:SM_ex_SCP}A and C show the base plate with one and two plate stiffeners per edge, respectively. Figure \ref{fig:SM_ex_SCP}D shows the corresponding simplified model where the adjacent plate stiffeners have been converted to edge moments. The resulting buckling strengths are also shown in table \ref{tab:Stiffener}. There is a discrepancy between the physical plate stiffeners compared to the simplified model. This is attributed to the planar stiffness of plates compared to the assumption of treating plate stiffeners as wide beams. However, if the plate stiffeners are split into narrower plate strips (see figure \ref{fig:SM_ex_SCP}B), the buckling strength converges to the simplified model results, as the strips become narrower (see figure \ref{fig:SM_ex_SCP}E). Note that the gap seen in the figures are enlarged for visual purposes, whereas they are much smaller in the numerical simulations.

In summary, the simplified model accounts for the lattice structure topology by converting rotational stiffness from adjacent members into moments. This is significantly cheaper than for example a full 3D Floquet-Bloch analysis, however still a more consistent approach than simple analytical column or plate buckling calculations.

\subsubsection{Material buckling strength estimation}
The simplified model is employed to estimate the material buckling strength of the investigated lattice structures subjected to a uni-axial load aligned with the Cartesian x-axis. 

\begin{figure}[!b]
  \centering
  \includegraphics[width=1\textwidth]{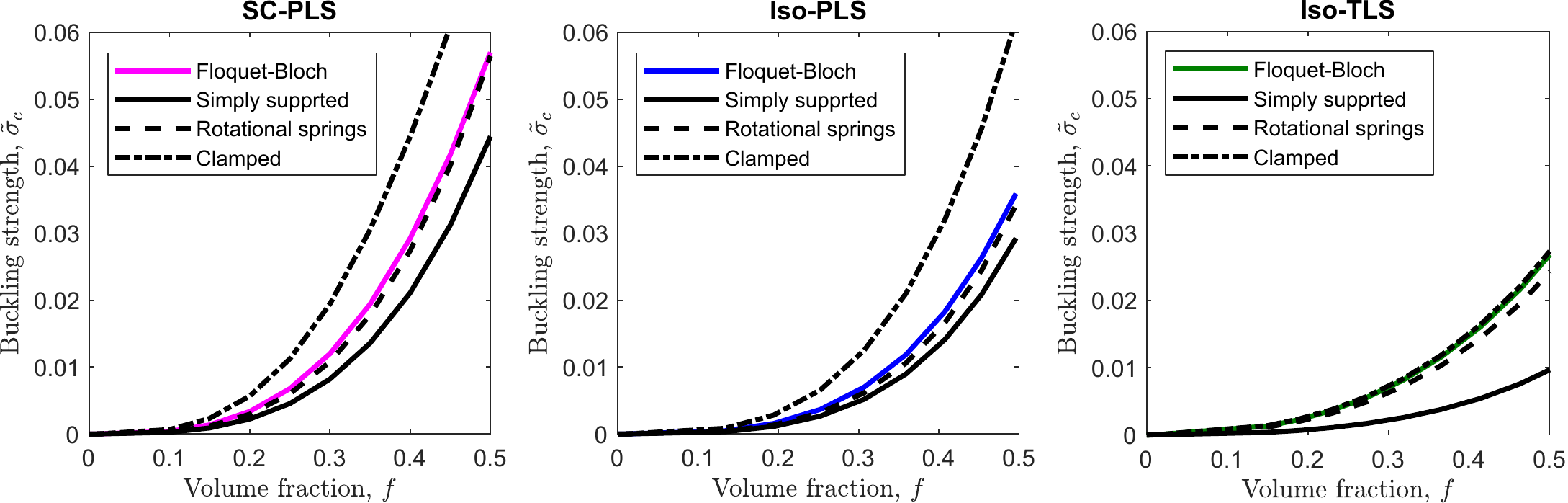}
\caption{Normalized buckling strength versus volume fraction for the uni-axial case of the load aligned with a principal axis, e.g. the Cartesian x-axis (see figure \ref{fig:lattices}). Simply supported, rotational springs and clamped boundary conditions are compared to full unit cell Floquet-Bloch results. A: SC-PLS. B: Iso-PLS. C: Iso-TLS. }
\label{fig:BC_study}
\end{figure}

Figure \ref{fig:BC_study} shows the impact on the buckling strength versus volume fractions under the uni-axial load along one principle axis when applying simply supported, rotational springs and clamped boundary conditions. These results are compared to the true buckling strength by using Floquet-Bloch wave theory on the unit cells from section \ref{sec:FloquetBloch}. For SC-PLS and Iso-PLS, it is clearly seen that the simply supported and clamped boundary conditions under- and overestimate the buckling strength, respectively. In contrast, the boundary conditions with rotational springs provide much improved estimations. In the case of Iso-TLS, the clamped boundary condition is appropriate due to the large number of truss members overlapping and thus stiffening the ends of trusses. However, for other load cases that triggers different critical buckling modes or TLS topologies with fewer overlapping trusses, this might not be true. Nevertheless, the boundary conditions with rotational springs provide a sufficiently accurate estimation of the true buckling strength. This observation promises that the effective buckling strength of the whole lattice structure can be predicted by the buckling behaviour of individual members. 

\begin{figure}[!t]
  \centering
  \includegraphics[width=0.9\textwidth]{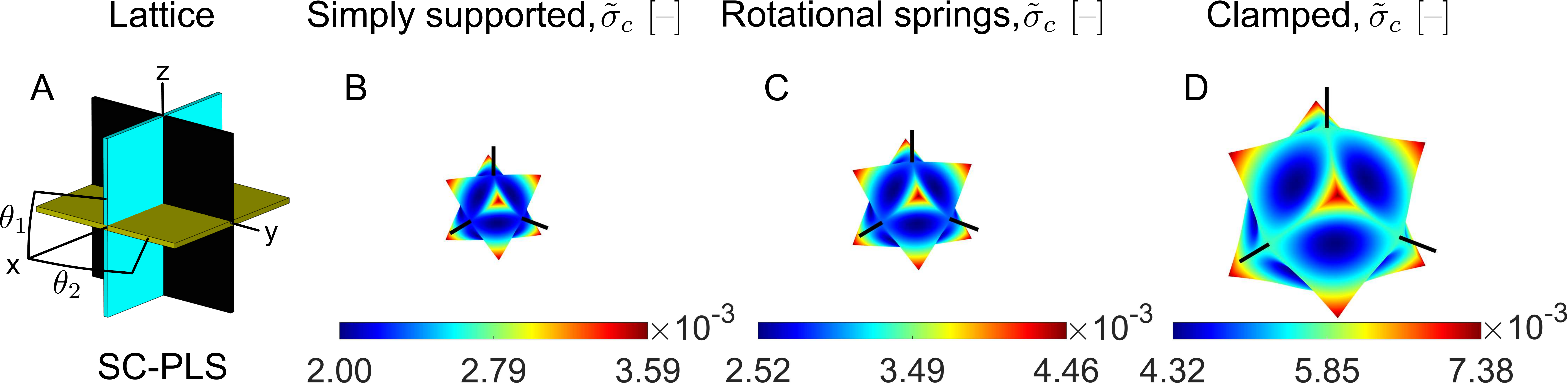}
\caption{Normalized buckling strength surfaces of SC-PLS subjected to different boundary conditions. A: Lattice typology with colored plate member sets and depiction of load orientation described by two Euler angles. B-D: Buckling strength surface from applying simply supported boundary conditions (B), rotational springs (C) and clamped boundary conditions (D)  to the plates. All surfaces are depicted with Cartesian coordinate system axes.}
\label{fig:SCP_surfaces}
\end{figure}

Figure \ref{fig:SCP_surfaces} shows the impact of the boundary conditions on the buckling strength surfaces for arbitrary directional uni-axial loads. The rotated uniaxial loads are obtained by $\boldsymbol{\sigma}_\text{rot}^0=\boldsymbol{R} \boldsymbol{\sigma}_\text{m}^0 \boldsymbol{R}^T$, where  $\boldsymbol{\sigma}_\text{m}^0$ is the matrix-formated uniaxial load of $\boldsymbol{\sigma}^0= \left[-1,0,0,0,0,0 \right]^T $ and  $\boldsymbol{R}$  is a rotation matrix containing two Euler angles $\theta_1$ and $\theta_2$ that span over $0$ to $\pi/2$ to cover the entire rotated stress (see figure \ref{fig:lattices}A) \cite{wang2020quasiperiodic}. The surfaces are described by two Euler angles and the associated buckling strength magnitudes describe the size. Interestingly, the surfaces do not have the exact same shape, which means that the effect of boundary conditions is not just a scaling of the strength. Specifically, it can be observed that when going from a Cartesian coordinate system axis to a near maximum strength angle, the path is flat for the simply supported case (Figure \ref{fig:SCP_surfaces}B), while it is curved for the clamped case (Figure \ref{fig:SCP_surfaces}D), highlighting the non-linear contributions from the adjacent member stiffnesses. 

In summary, the simplified model yields sufficiently accurate estimations of the material buckling strength with a significant improvement compared to the simply supported and clamped boundary conditions. Furthermore, it is shown that the simply supported and clamped boundary conditions may only be used as lower and upper bounds.

%%%%%%%%%%%%%%%%%%%%%%%%%%%%%%%%%%%%%%%%%%%%%%%%%%%%%%%%%%%%%

\section{Results and discussion} \label{sec:Results}
In this section, we employ the simplified model to investigate the performance of the lattice structures under rotated uni-axial loads.
%%%%%%%%%%%%%%%%%%%%%%%%%%%%%%%%%%%%%%%%%%%%%%%%%%%%%%%%%%%%%%
\subsection{Strength surfaces}

\begin{figure}[!t]
  \centering
  % include first image
  \includegraphics[width=\textwidth]{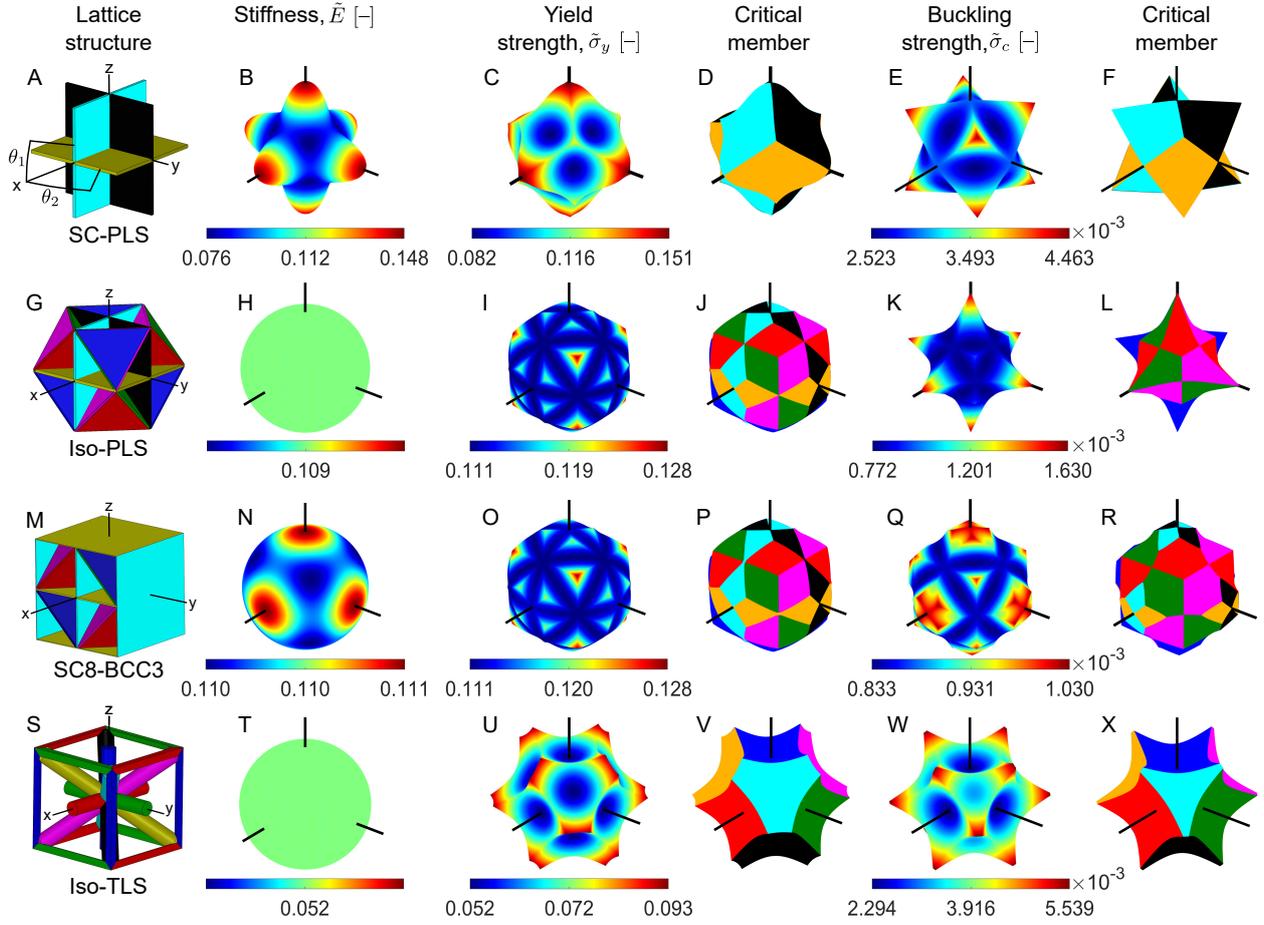}
\caption{Normalized directional stiffness and strength surfaces of the lattice structures at volume fraction $f\approx 0.2$. A, G, M and S: Lattice topologies with colored plate or truss member sets and depiction of load orientation described by two Euler angles. The front black plate is hidden for the SC8-BCC3.  B, H, N and T: Directional stiffness surfaces. C, I, O and U: Directional yield strength surfaces. D, J, P and V: colored member set that fails due to yielding under given load direction. E, K, Q and W: Directional buckling strength surface. F, L, R and X: colored member set that fails due to buckling under given load direction. All surfaces are depicted with Cartesian coordinate system axes.}
\label{fig:Surfaces}
\end{figure}

Figure \ref{fig:Surfaces} shows the directional stiffness and strength surfaces of the SC-PLS, Iso-PLS, Iso-TLS, and an improved buckling strength isotropic configuration of the Iso-PLS, which will be discussed later. The two Euler angles that are used to describe the load direction are shown in figure \ref{fig:Surfaces}A, where the principal axes in the Cartesian coordinate system are shown on all the lattices and surfaces.

The SC-PLS has the highest and lowest directional stiffness when the load is aligned with two SC plates and is furthest away from all plates, respectively (figure \ref{fig:Surfaces}B). Correspondingly, the highest yield strength arises when the load is aligned with two plates (figure \ref{fig:Surfaces}C). The highest yield strength angles can intuitively be seen where the critical plate that has the lowest yield strength for a given angle is at an intersection of different plate sets (figure \ref{fig:Surfaces}D). For comparison, the lowest yield strength arises when the load is aligned with any single plate. In order to get a comparison of the strength anisotropy of the lattice structures, a ratio of the highest to the lowest strength is calculated by $\tilde{\sigma}_\text{y,max}/\tilde{\sigma}_\text{y,min}=1.84$. Furthermore, the highest buckling strength angle arises when the stresses are evenly distributed among all the plates (figure \ref{fig:Surfaces}E), where the buckling strength ratio is $\tilde{\sigma}_\text{c,max}/\tilde{\sigma}_\text{c,min}=1.77$. Once again, the angles for the highest strength are located at intersections between critical plate sets (figure \ref{fig:Surfaces}F). 

The Iso-PLS has isotropic directional stiffness, visualized by a perfect sphere (figure \ref{fig:Surfaces}H). Unfortunately, isotropic stiffness does not necessarily result in isotropic strength. Nevertheless, the yield strength surface is close to being isotropic (figure \ref{fig:Surfaces}I), where it has $\tilde{\sigma}_\text{y,max}/\tilde{\sigma}_\text{y,min}=1.16$. The lowest and highest yield strength appear when the load is aligned and furthest away from the plate sets, respectively. The highest yield strength angles can also be seen at the intersection points by three plate sets (figure \ref{fig:Surfaces}J). This excludes intersection points where the load is aligned with any single plate set, e.g. at $\theta_1=0\deg$ and $\theta_2=45\deg$. In contrast to the near isotropic yield strength surface, the buckling strength surface is much more anisotropic (figure \ref{fig:Surfaces}K), which yields $\tilde{\sigma}_\text{c,max}/\tilde{\sigma}_\text{c,min}=2.11$. The highest buckling strength is obtained when the load is aligned with two of the SC plate sets (the black, yellow, or cyan plate sets). However, the logical presumption that the SC plates would buckle is not true. Instead, studying the associated critical plate visualization (figure \ref{fig:Surfaces}L), reveals that only the BCC plates are susceptible to buckling failure. The reason is that the BCC plates (the red, green, blue, and magenta plate sets) are significantly thinner and hence the highest buckling strength is located at the angle furthest away from the BCC plates, i.e. aligned with SC plates which also helps by carrying the majority of the stress. Correspondingly, the lowest buckling strength is when the load is aligned with any single BCC plate set.

The Iso-TLS also has isotropic stiffness (figure \ref{fig:Surfaces}T), but its yield strength surface is more anisotropic than the Iso-PLS (figure \ref{fig:Surfaces}U). This is because trusses only provide axial stiffness while plates offer planar stiffness. Nevertheless, its highest yield strength is also located at the load angle furthest away from any truss set directions, where $\tilde{\sigma}_\text{y,max}/\tilde{\sigma}_\text{y,min}=1.79$. The highest yield strength can also be seen by studying the critical truss (figure \ref{fig:Surfaces}V), which is again at the intersection of three different truss sets. Furthermore, its buckling strength surface (figure \ref{fig:Surfaces}W) resembles the yield strength surface, with $\tilde{\sigma}_\text{c,max}/\tilde{\sigma}_\text{c,min}=2.41$. Hence, the same conclusions can be drawn for the critical buckling truss (figure \ref{fig:Surfaces}X). The difference in the surface shapes is due to the BCC trusses (the black, yellow, cyan, and magenta truss sets), being less slender and hence having higher resistance to buckling.

\subsubsection{Improved buckling strength isotropic Iso-PLS configuration}\label{sec:ImprovedIsoPLS}
Based on the knowledge gained from the simplified model and the critical member plots, the isotropy of the buckling strength surface of Iso-PLS can be improved. Specifically, this can be achieved by splitting single SC plates into two parts by keeping one half of the thickness in the middle and distributing the other half to the outer faces of the unit cell, which creates a boundary box (see figure \ref{fig:Surfaces}M, where the front black plate is hidden). Following the naming convention from \cite{tancogne2018elastically} makes the original Iso-PLS an SC-BCC3 PLS and the new configuration an SC8-BCC3 PLS. This new configuration only consists of the theoretically more isotropic tetrahedral holes, whereas the original Iso-PLS has extra octahedral holes when tessellated through space (e.g. the closest blue plate in figure \ref{fig:Surfaces}G). This new configuration stiffens all the BCC plates equally, while not changing the stiffness or yield strength surfaces, as the same amount of volume is retained in all directions. By simply splitting the SC plates, there is a change in the amount of overlapping volume. In order to recover perfect stiffness isotropy, the thickness ratio between the SC and BCC plates has to be slightly adjusted (figure \ref{fig:Surfaces}N). The shapes of the yield strength and critical member surfaces of the new lattice structure are unchanged (figure \ref{fig:Surfaces}O and P). In contrast, its buckling strength surface (figure \ref{fig:Surfaces}Q), is more isotropic and resembles the yield strength surface. The lowest and highest strengths remain at the same directions as above, but the buckling strength ratio is now reduced to $\tilde{\sigma}_\text{c,max}/\tilde{\sigma}_\text{c,min}=1.24$. Note that, recovering perfect stiffness isotropy would decrease and increase the thicknesses of SC and BCC plates, respectively, resulting in a lower buckling strength ratio. Lastly, the resemblance of the yield and buckling strength surfaces can also be seen by studying the critical plates (figure \ref{fig:Surfaces}R).

\subsubsection{Summary}
Isotropic stiffness does not necessarily result in isotropic strength. Nevertheless, Iso-TLS and SC8-BCC3 have yield strength surfaces that resemble their buckling strength surfaces. In contrast, SC-PLS and Iso-PLS have vastly different strength surfaces. Similar and more isotropic strength surfaces result in more predictable failure mechanisms based on the volume fractions and the base material, whereas dissimilar surfaces also depend on load orientations. In general, the highest yield and buckling strength of the lattice structures arise when the stresses are equally distributed between the members, for instance when the load is furthest away from the members or when a majority of members are aligned with the load direction. Furthermore, low slenderness of a member also provides high buckling strength. Interestingly, Iso-PLS has higher stiffness and yield strength for all load directions than Iso-TLS, but conversely, Iso-TLS has higher buckling strength for all load directions.

\subsection{Verification of buckling strength surfaces}

\begin{table}[!t]
\centering
{\renewcommand{\arraystretch}{1.2} %<- modify value to suit your needs
\resizebox{\textwidth}{!}{%
\begin{tabular}{|c|c|c|c|c|c|c|}
\hline
Lattice &
  $\theta_1 \deg$ &
  $\theta_2\deg$ &
  Floquet-Bloch &
  \begin{tabular}[c]{@{}c@{}}Simply supported\\ (error \%)\end{tabular} &
  \begin{tabular}[c]{@{}c@{}}Rotational springs\\ (error \%)\end{tabular} &
  \begin{tabular}[c]{@{}c@{}}Clamped\\ (error \%)\end{tabular} \\ \hline
\multirow{5}{*}{SC-PLS} &
  0 &
  0 &
  3.36E-3 &
  \begin{tabular}[c]{@{}c@{}}2.22E-3\\ (33.84)\end{tabular} &
  \begin{tabular}[c]{@{}c@{}}2.94E-3\\ (12.51)\end{tabular} &
  \begin{tabular}[c]{@{}c@{}}5.66E-3\\ (-68.37)\end{tabular} \\ \cline{2-7} 
 &
  0 &
  45 &
  3.39E-3 &
  \begin{tabular}[c]{@{}c@{}}2.00E-3\\ (41.10)\end{tabular} &
  \begin{tabular}[c]{@{}c@{}}2.52E-3\\ (25.56)\end{tabular} &
  \begin{tabular}[c]{@{}c@{}}4.32E-3\\ (-27.52)\end{tabular} \\ \cline{2-7} 
 &
  35.26 &
  45 &
  5.12E-3 &
  \begin{tabular}[c]{@{}c@{}}3.59E-3\\ (29.96)\end{tabular} &
  \begin{tabular}[c]{@{}c@{}}4.47E-3\\ (12.86)\end{tabular} &
  \begin{tabular}[c]{@{}c@{}}7.38E-3\\ (-44.10)\end{tabular} \\ \hline
\multirow{5}{*}{Iso-PLS} &
  0 &
  0 &
  1.91E-3 &
  \begin{tabular}[c]{@{}c@{}}1.37E-3\\ (28.22)\end{tabular} &
  \begin{tabular}[c]{@{}c@{}}1.63E-3\\ (14.66)\end{tabular} &
  \begin{tabular}[c]{@{}c@{}}3.42E-3\\ (-79.30)\end{tabular} \\ \cline{2-7} 
 &
  24.09 &
  26.57 &
  1.32E-3 &
  \begin{tabular}[c]{@{}c@{}}6.43E-4\\ (51.27)\end{tabular} &
  \begin{tabular}[c]{@{}c@{}}7.72E-4\\ (41.50)\end{tabular} &
  \begin{tabular}[c]{@{}c@{}}1.67E-3\\ (-26.85)\end{tabular} \\ \cline{2-7} 
 &
  35.26 &
  45 &
  1.40E-3 &
  \begin{tabular}[c]{@{}c@{}}7.83E-4\\ (44.09)\end{tabular} &
  \begin{tabular}[c]{@{}c@{}}9.38E-4\\ (32.99)\end{tabular} &
  \begin{tabular}[c]{@{}c@{}}2.02E-3\\ (-44.41)\end{tabular} \\ \hline
\multirow{5}{*}{SC8-BCC3} &
  0 &
  0 &
  1.53E-3 &
  \begin{tabular}[c]{@{}c@{}}7.84E-4\\ (48.58)\end{tabular} &
  \begin{tabular}[c]{@{}c@{}}1.03E-3\\ (32.78)\end{tabular} &
  \begin{tabular}[c]{@{}c@{}}2.07E-3\\ (-35.77)\end{tabular} \\ \cline{2-7} 
 &
  24.09 &
  26.57 &
  1.36E-3 &
  \begin{tabular}[c]{@{}c@{}}6.47E-4\\ (52.57)\end{tabular} &
  \begin{tabular}[c]{@{}c@{}}8.33E-4\\ (38.91)\end{tabular} &
  \begin{tabular}[c]{@{}c@{}}1.68E-3\\ (-23.47)\end{tabular} \\ \cline{2-7} 
 &
  35.26 &
  45 &
  1.55E-3 &
  \begin{tabular}[c]{@{}c@{}}7.87E-4\\ (49.36)\end{tabular} &
  \begin{tabular}[c]{@{}c@{}}1.01E-3\\ (34.95)\end{tabular} &
  \begin{tabular}[c]{@{}c@{}}2.03E-3\\ (-30.72)\end{tabular} \\ \hline
\multirow{5}{*}{Iso-TLS} &
  0 &
  0 &
  2.59E-3 &
  \begin{tabular}[c]{@{}c@{}}7.56E-4\\ (70.83)\end{tabular} &
  \begin{tabular}[c]{@{}c@{}}2.30E-3\\ (11.34)\end{tabular} &
  \begin{tabular}[c]{@{}c@{}}2.68E-3\\ (-3.30)\end{tabular} \\ \cline{2-7} 
 &
  0 &
  45 &
  4.37E-3 &
  \begin{tabular}[c]{@{}c@{}}2.28E-3\\ (47.92)\end{tabular} &
  \begin{tabular}[c]{@{}c@{}}5.54E-3\\ (-26.78)\end{tabular} &
  \begin{tabular}[c]{@{}c@{}}7.43E-3\\ (-69.93)\end{tabular} \\ \cline{2-7} 
 &
  35.26 &
  45 &
  3.54E-3 &
  \begin{tabular}[c]{@{}c@{}}1.32E-3\\ (62.74)\end{tabular} &
  \begin{tabular}[c]{@{}c@{}}3.21E-3\\ (9.30)\end{tabular} &
  \begin{tabular}[c]{@{}c@{}}4.30E-3\\ (-21.57)\end{tabular} \\ \hline
\end{tabular}%
}}
\caption{Comparison of buckling estimates to unit cell Floquet-Bloch wave analysis. Two angles of interest per lattice are compared to verify the simplified model with the addition of the load oriented along principal axes from figure \ref{fig:BC_study}.}
\label{tab:Comparison}
\end{table}

The buckling strength surfaces from the simplified model are verified by investigating two load angles of interest per lattice structure, besides the load aligned with a principal axis (figure \ref{fig:BC_study}). The verification is performed by using FE-based Floquet-Bloch wave analysis on the single unit cells following the approach from section \ref{sec:FloquetBloch}. The comparison is shown in table \ref{tab:Comparison} with the addition of results from simply supported and clamped boundary conditions. The error is calculated as $\left(\tilde{\sigma}_\text{c,FB}-\tilde{\sigma}_\text{c,SM}\right)/\tilde{\sigma}_\text{c,FB}\cdot 100\%$, where a positive or negative value implies under- or overestimation, respectively. 

For the SC-PLS, the investigated angles are the minimum strength angle where the load is aligned with only one SC plate as well as the maximum strength angle where the load is furthest away from all plates. For the Iso-PLS and SC8-BCC3, the investigated angles are the minimum strength angle where the load is aligned with only the green BCC plates and the high strength angle where the load is perpendicular to the blue BCC plates. For the Iso-TLS, the investigated angles are the high strength angle where the load is aligned with the cyan BCC truss set and the maximum strength angle where the load is furthest away from all truss sets. For the maximum strength angle, the simplified model overestimates the buckling strength by $-26.78\%$, while the simply supported and clamped boundary conditions lead to underestimation of $47.92\%$ and overestimation of $-69.93\%$, respectively. Here, the reason for the overestimation of the simplified model is that the applied stress triggers a critical buckling mode which has rotation at the ends of the trusses and hence clamped or close to clamped boundary conditions result in an overestimation.

We remark that the reported errors are from the extreme cases of maximum and minimum strengths often associated with peaks and dips in the buckling surfaces and hence extra susceptible to errors. The average errors are much smaller. The errors resulting from the simplified model are partly from the simplification of the effective length, which is especially pronounced for the isotropic lattice structures as they have non-perpendicular overlapping members and thus the shortening effects are greater. Furthermore, the inter-connectivity between the stiffeners of the PLSs is not considered, which again is especially pronounced for the isotropic PLSs as they have many overlapping members.

Despite obvious errors, the simplified model provides an overall much better estimation compared to the simply supported and clamped boundary conditions. On the other hand, it is infeasible to create microstructural buckling strength surfaces based on numerical results on single unit cell analyses. For example, for the investigated lattice structures (figure \ref{fig:lattices}) subjected to complex load cases where the Brillouin zone cannot be reduced to the boundaries of a polyhedron and at least needs a crude sweep over $9\cdot 9 \cdot 9$ wavevectors $k_j \in \left[-\pi,\ \pi\right]$, the average computational time using $24$ CPU's is approximately $100$ hours in COMSOL per load case per lattice \cite{andersen2020competition}. In contrast, the computational time of the simplified model is approximately $5$ seconds per member set. Hence, the simplified model, supported by some unit cell analysis sanity checks, is a viable solution to investigate the strengths and weaknesses of lattice structures.

\section{Conclusions} \label{sec:conc}
We propose a simplified model for predicting the strength of stretch-dominated TLSs and PLSs, which yields good accuracy compared to very costly Floquet-Bloch wave analysis. The model is used to create directional buckling strength surfaces showing the strength anisotropy of the considered lattice structures. The findings reveal that isotropic stiffness does not necessarily result in isotropic (both yield and buckling) strength and that yield and buckling strength surfaces can also be widely different. 

It is by now known that PLSs have higher stiffness and yield strength, while TLSs have higher buckling strength. However, here we show that the PLSs can have vastly different directional buckling strength surfaces compared to their yield strength surfaces. For specific combinations of the base material and volume fractions, either buckling or yielding could dominate the critical failure mode depending on the load directions. In contrast, (isotropic) TLSs have very similar buckling and yield strength surfaces, resulting in more predictable failure modes, which are almost purely based on the base material and volume fractions. 

We use the knowledge gained from the simplified model to create a new configuration of the isotropic plate lattice structure to improve the isotropy of the buckling strength surface. Furthermore, the yield and buckling strength surfaces are also very similar for this configuration.

Future studies include investigation of the sensitivity and determination of the worst-case scenario of lattices subjected to arbitrary loads following an approach such as \cite{wang2019simple}. Furthermore, it is also interesting to investigate how the simplified model compares to results from finite lattice studies \cite{wang2020numerical} and specifically when utilized as infill in macrostructures \cite{clausen2016exploiting}. Finally, the simplified model paves the way for efficiently accounting for local buckling failure in multiscale topology optimization \cite{groen2020homogenization}. 

\section*{Acknowledgements}
We acknowledge the financial support from the Villum Investigator Project InnoTop.

%%%%%%%%%%%%%%%%%%%%%%%%%%%%%%%%%%%%%%%%%%%%%%%%%%%%%%%%%%%%%

\bibliographystyle{elsarticle-num-names}
\bibliography{References}
 	
\end{document}